\documentclass[12pt,a4paper]{article}
\usepackage{graphicx,amsmath,amsfonts}
\def \ba{\begin{eqnarray}}\def\ea{\end{eqnarray}}
\def\bc{\begin{center}}\def\ec{\end{center}}
\def\nn{\nonumber\\}
\title{\huge \bf Electromagnetic corrections to final state
interactions in  $K\to 3\pi$ decays}
\author{\bf S.R.Gevorkyan\footnote{On leave of absence from
Yerevan Physics Institute}, A.V.Tarasov,
O.O.Voskresenskaya\footnote{On leave of absence from Siberian
Physical Technical Institute} }
\date{}
\begin{document}
\maketitle \bc Joint Institute for Nuclear Research, 141980 Dubna,
Russia \ec
\begin{abstract}
  The final state interactions of pions in decays
$K^\pm\to\pi^\pm\pi^0\pi^0$  are  considered using the  methods of
quantum mechanics. We show how to incorporate the electromagnetic
effects in the amplitudes of these decays and to work out the
relevant expressions valid  above and below the two charged pions
production threshold $M_c=2m$. The electromagnetic corrections are
given as evaluated in a potential model.
\end{abstract}
During the last years essential progress has been achieved in
$\pi\pi$ scattering lengths determination from experimental
data~\cite{A,P,B}.The precise knowledge of these quantities is an
important task,since at present the Chiral Perturbation Theory
(ChPT)  predicts their  values with very high accuracy~\cite{CGL}.\\
The high quality data on $K^\pm\to \pi^\pm\pi^0\pi^0$ decays have
been obtained recently in the NA48/2 experiment at CERN
SPS~\cite{B}.Dependence of the decay rate on the invariant mass of
neutral pions $M^2=(p_1+p_2)^2$ reveals a prominent anomaly (cusp)
at the threshold, relevant to the production of two charged
pions $M_c^2=4m^2$.\\
The theoretical investigations of the decays \ba  K^\pm\to \pi^\pm
\pi^0\pi^0,\ea \ba K^\pm\to \pi^\pm \pi^+\pi^-\ea have been carried
out for many years~\cite{G1,G2,AA}.New experimental data of high
accuracy ~\cite{B} lead  to great activity on this issue~\cite{C,CI,Gass}.\\
As was explained by N.~Cabibbo \cite{C},the cusp in the experimental
decay distribution is a result of the charge exchange scattering
process $\pi^+\pi^-\to \pi^0\pi^0$ ~in the decay (2).He proposed a
simple re-scattering model,in which the amplitude of decay (1)
consists of two terms \ba T=T_0+2ika_xT_+\ea where $T_0,~T_+$ are
the ``unperturbed'' amplitudes for decays (1) and (2) respectively
and $k=\frac{1}{2}\sqrt{M^2-4m^2}$ is the momentum of the charged
pion. The second term in (3) is proportional to the difference of
scattering lengths $a_x=(a_0-a_2)/3$ and flips from dispersive to
absorptive at the threshold $M_c$. As a result  the decay
probability under the threshold linearly depends  on $a_x$
allowing  this difference to be extracted from experimental data with high accuracy.\\
The next important step  was  done in ~\cite{CI}, where the
amplitude T was obtained accounting the second order in scattering
lengths terms using analyticity and unitarity of the S matrix. The
results of~\cite{CI} were supported by rigorous
approaches~\cite{Gass,GPS}, whose authors investigated  the decay
(1) in the framework of effective field theory and ChPT and
calculate all re-scattering processes in two-loop approximation. The
results from~\cite{CI} were used in the fit of experimental
data~\cite{B},allowing  extraction of the difference $a_0-a_2$ with high accuracy.\\
Nevertheless the decay rate behavior near threshold cannot be
provide solely by strong interaction of pions in the final state.As
was widely discussed (see e.g. ~\cite{I,G}), the tiny discrepancy
between theoretical predictions and experimental data in the
vicinity of threshold ~\cite{B} is a result of disregarding
electromagnetic effects in final state interaction. In view of
importance of the knowledge of the scattering lengths with the most
possible accuracy, the consideration of electromagnetic
corrections in decay (1) becomes a pressing  issue.\\
Later on we discuss the problem of Coulomb interaction among charged
pions  using methods of non-relativistic quantum mechanics, which
are completely suitable for the considered case\footnote{The effects
of radiation of real photons are beyond the scope of our
consideration and will be treated elsewhere}. We obtain compact
expressions for the amplitude \footnote{As the $K^-$ decays are
counterparts to the $K^+$ decays,they are not treated separately.}
of  decay (1) with regard to the electromagnetic corrections, which
are valid  below and above the charged pions production threshold $M_c=2m$.\\
Leaving the strict derivation  for a separate publication,we shortly
discuss how to involve the electromagnetic effects in the considered
problem and relevant modifications, which have to be done
in the amplitude of decay  (1).\\
With the methods of non-relativistic quantum mechanics it can be
shown that the result of N.~Cabibbo~\cite{C} can be generalized
accounting the $\pi\pi$ scattering to all orders in scattering
lengths \ba T&=&T_0+2ikf_xT_+,\quad f_x=a_x/D,\nn
D&=&(1-ik_1R_{11})(1-ik_2R_{22})+k_1k_2R_{12}^2\ea Here
$k_1=\frac{1}{2}\sqrt{M^2-4m_0^2};k_2=k=\frac{1}{2}\sqrt{M^2-4m^2}$
are the momenta of neutral and charged pions  respectively. The
elements of the R matrix are real and can be expressed in isospin
symmetry limit through the combinations of the scattering
lengths~\cite{CI,Gass} $a_x=(a_2-a_0)/3;a_{00}=(a_0+2a_2)/3;
a_\pm=(2a_0+a_2)/6$ corresponding to inelastic and elastic pion-pion
scattering as \ba R_{12}=\sqrt{2}a_x;~~R_{11}=a_{00};~~R_{22}=2a_\pm
\ea The replacement $a_x\to f_x$ has small numerical impact on the
results of the previous calculations done according ~\cite{CI,Gass}
in the dominant part of phase space, but, as we will see later,is
very crucial for inclusion of the electromagnetic interactions under
the threshold,where formation of bound states ($\pi^+\pi^-$ atoms) take place.\\
The next step of our prescription is inclusion of electromagnetic
effects in expression (4). The general receipt is known for many
years (see for instance the textbook~\cite{BZP}) and implies
replacement of charged pion momenta k  by a logarithmic derivative
of the pion wave function in the Coulomb potential at the boundary
of the strong field  $r_0$ i.e. \ba ik\to \tau=\frac{d\log[G_0(k
r)+iF_0(k r)]}{dr}\biggl |_{r=r_0}\ea
Here $F_0,G_0$ are the regular and irregular solutions of the Coulomb problem.\\
In the region $ kr_0\ll 1$ where the electromagnetic effects can be
significant the above replacement gives \ba \tau &=& ik-\alpha
m\left[\log(-2ikr_0)+2C+\psi\left(1-\frac{im\alpha}{2k}\right)\right]\nn
&=&Re(\tau)+i Im(\tau)\nn Re(\tau)&=&-\alpha m\left[\log(2kr_0)+2C+
Re\,\psi\left(1-\frac{im\alpha}{2k}\right)\right],\nn Im(\tau)&=&
ikA^2,\quad
A=\exp{\left(\frac{\pi\xi}{2}\right)}|\Gamma(1+i\xi)|,\quad
\xi=\frac{\alpha m}{2k}\ea where $C=0.577$, $\alpha=1/137$ are the
Euler and fine structure constants, whereas $\psi$ is the digamma function.\\
To go under the threshold, it is enough  to make the common
replacement $k\to i\kappa $ in the above expression \ba
\tau=-\kappa-\alpha m\left[\log(2\kappa
r_0)+2C+\psi\left(1-\frac{m\alpha}{2\kappa}\right)\right]\ea At
$\kappa_n=\alpha m/(2n)$, where n is an integer, $\tau$ goes to
infinity, which corresponds to Coulombic bound states in the
considered approach. On the other hand, the product $\kappa f_x$
defining the amplitude behavior  under the threshold, remains finite
due to dependence on $\tau$ in the denominator D in expression (4).
This explains why electromagnetic effects can be included only after
summing up all terms of the infinite series in the perturbation expansion.\\
The product $\kappa f_x$ possesses  a resonance structure placed at
the positions \ba
M_n&=&2m-\frac{\bar\kappa_n^2}{m},~~~\bar\kappa_n=\frac{\alpha
m}{2(n-\delta)},\nn \delta &=&\frac{1}{\pi}\arctan \Delta,~~~
\Delta=\alpha
m\left[a_{22}-\frac{k_1^2a_{11}a^2_{12}}{1+k_1^2a^2_{11}}\right]\ea
with the relevant width \ba \Gamma_n=\frac{4\pi
k_1a_{12}^2\bar\kappa_n^3}{m(1+k_1^2a^2_{11})}.\ea The physical
reason of resonance origin is transparent. Due to the charge
exchange  process  $\pi^+\pi^-\to \pi^0\pi^0 $ the Coulombic bound
states of the  $\pi^+\pi^-$ system ($A_{2\pi}$ atoms) becomes
unstable\footnote{We do not discuss here the  instability of excited
states caused by their transition to the ground state, as this
effect is very small and can safely be neglected.}.\\
The considered effect of the creation  of $A_{2\pi}$ atoms  in decay
(1) is not the only  contribution from electromagnetic interaction of pions.\\
Outside the resonance region the Coulomb interaction leads to the
essential difference between the $\tau$ values calculated with
electromagnetic corrections and without them. In particular, the
nonzero contribution of the Coulomb corrections to the $Re(\tau)$
above the threshold leads to the interference term in the decay rate
provided by ``direct'' and ``charge-exchange'' contributions from
(4). Thus above the threshold the interference is nonzero  even at
the lowest order in scattering lengths, unlike the original
approach proposed by N.~Cabibbo~\cite{C}.\\
Further improvement of the theory consists in taking account of
final state interactions in the ``direct'' term from (4). This can
be done by simple substitution \ba T_0\to T_0(1+ik_1f_{00})\ea
where $f_{00}$ is the  amplitude of $\pi^0\pi^0$ scattering.\\
It can be shown that \ba 1+ik_1f_{00}=\frac{1-\tau
a_{22}}{(1-ik_1a_{11})(1-\tau a_{22})-ik_1\tau a_{12}^2} \ea These
higher order corrections to "direct" term are numerically small, but
taking in account the precision of experimental data, have to be
included in the fit procedure.\\
 To estimate the contribution from electromagnetic effects to the decay
 rate of process (1), we introduce the ratio $R(\%)=\frac{|T_c|^2-|T|^2}{|T|^2}$
 where the amplitude T is given by expression (4), while the amplitude $T_c$
taking  account of electromagnetic effects is given by expressions
(4,11,12) with relevant modifications discussed  above. The dotted
line in Fig.1 shows the contribution of electromagnetic effects
without bound state  corrections. The dashed line represents the
same quantity, but with the corresponding  average with the gaussian
distribution~\cite{B} and the expected mass resolution near the
threshold $r.m.s.=0.56$ MeV. The solid line gives the contribution
of all electromagnetic effects (bound states included) averaged as
in the previous case with the gaussian distribution. From this plot
one concludes that the essential contribution to the decay rate in
the vicinity of the threshold
comes from the electromagnetic interactions that do not lead to bound states.\\
 The developed  approach allows one to take into account electromagnetic
effects in  decay (1) and to estimate their impact on decay rate of
the process under consideration.\\
We are grateful to V.~Kekelidze and J.~Manjavidze who draw our
attention to the problem and supported during  the work. We would
also like to thank D.~Madigozhin for many stimulating and useful
discussions.

\newpage

\begin{figure}[ht]
\begin{center}
\includegraphics[scale=0.7]{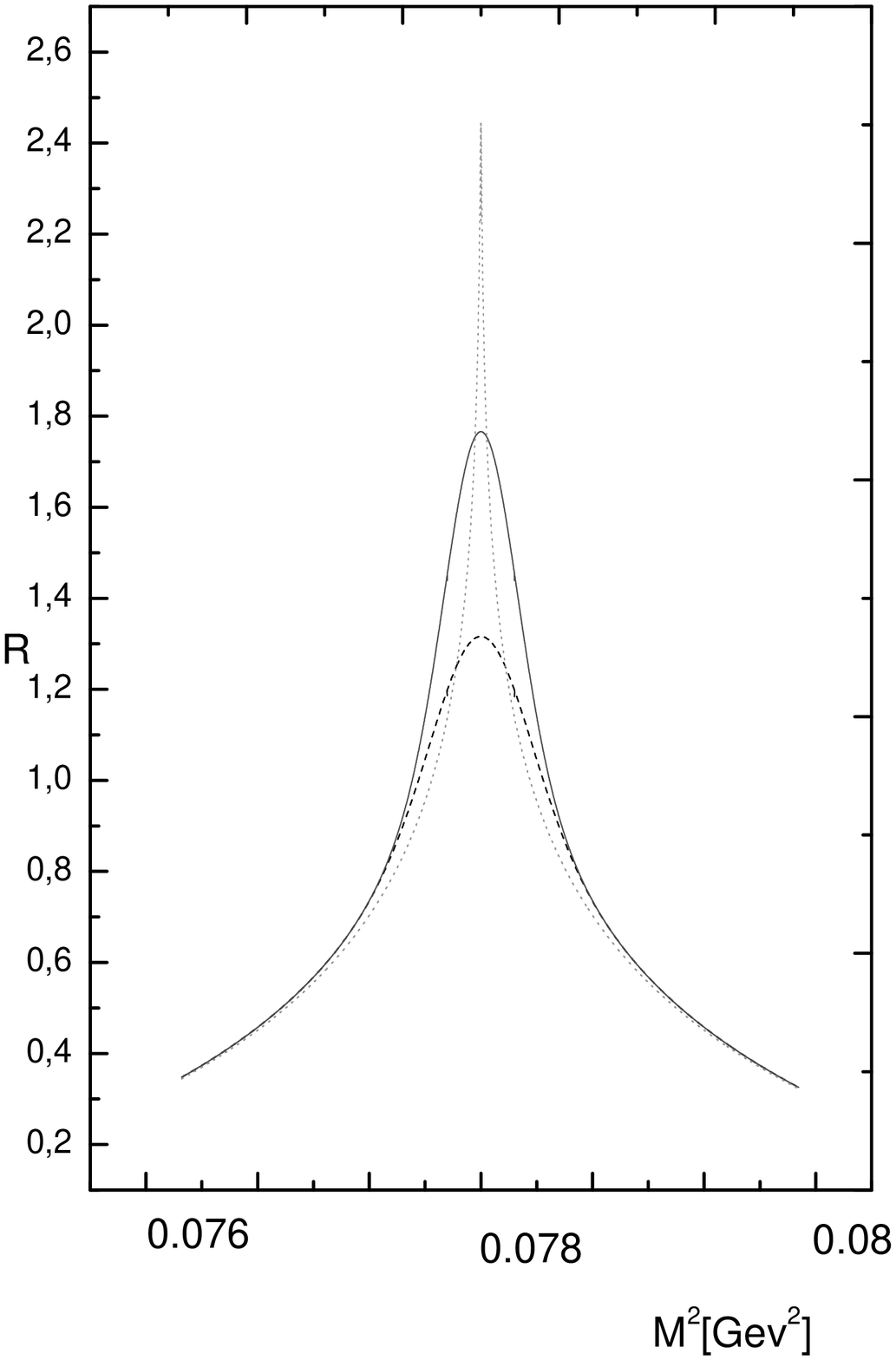}
\caption{Electromagnetic corrections contribution to decay rate as a
function of invariant mass of neutral pions.}
\end{center}
\end{figure}

\end{document}